\documentclass[12pt,a4paper,dvips]{article}
\usepackage{epsfig,wrapfig,times,mathptm} 
\renewcommand{\section}[1]{\vspace{6pt} \noindent\mbox{#1} \newline \noindent}
\renewcommand{\subsection}[1]{\vspace{6pt} \noindent\mbox{\underline{#1}} 
\newline \noindent}
\renewcommand{\subsubsection}[1]{\vspace{6pt} \noindent\mbox{\underline{#1}}
\noindent}

\newfont{\sansb}{cmssbx10}
\newfont{\sans}{cmss10}

\begin{document}
{\small OG 4.1.1 \vspace{-24pt}\\}     
{\center  \Large SPECTRUM OF TEV GAMMA RAYS FROM THE CRAB NEBULA
\vspace{6pt}\\}
D.A.Carter-Lewis$^1$, 
S.Biller$^2$, 
P.J.Boyle$^3$, 
J.H.Buckley$^4$, 
A.Burdett$^2$, 
J.Bussons Gordo$^3$,\\ 
M.A.Catanese$^1$, 
M.F.Cawley$^5$, 
D.J.Fegan$^3$,
J.P.Finley$^6$, 
J.A.Gaidos$^6$, \\
A.M.Hillas$^2$, 
F.Krennrich$^1$, 
R.C.Lamb$^7$, 
R.W.Lessard$^6$, 
C.Masterson$^3$, 
J.E.McEnery$^3$,\\ 
G.Mohanty$^1$, 
J.Quinn$^3$, 
A.J.Rodgers$^2$, 
H.J.Rose$^2$, 
F.W.Samuelson$^1$, 
G.H.Sembroski$^6$, \\
R.Srinivasan$^6$, 
T.C.Weekes$^4$, 
M.West,$^2$, 
J.A.Zweerink$^1$
\vspace{6pt}\\
{\it $^{1}$ Iowa State University, U.S.A.}\\
{\it $^{2}$ University of Leeds, United Kingdom.}\\
{\it $^{3}$ University College, Dublin, Ireland}\\
{\it $^{4}$ Whipple Observatory, Harvard-Smithsonian CfA, U.S.A.}\\
{\it $^{5}$ St.Patrick's College, Maynooth, Ireland}\\
{\it $^{6}$ Purdue University, U.S.A.}\\
{\it $^{7}$ Space Radiation Lab, Caltech, U.S.A.}
\vspace{12pt}
{\center ABSTRACT\\} The Crab Nebula has become established as the
standard candle for TeV gamma-ray astronomy using the atmospheric
Cherenkov technique.  No evidence for variability has been seen. The
spectrum of gamma rays from the Crab Nebula has been measured in the
energy range 500 GeV to 8 TeV at the Whipple Observatory by the
atmospheric Cherenkov imaging technique. Two methods of analysis
involving independent Monte Carlo simulations and two databases of
observations (1988-89 and 1995-96) were used and gave close
agreement. Using the complete spectrum of the Crab Nebula, the
spectrum of relativistic electrons is deduced and the spectrum of the
resulting inverse Compton gamma-ray emission is in good agreement with
the measured spectrum if the ambient magnetic field is about 25-30 nT.

\setlength{\parindent}{1cm}
\section{1. INTRODUCTION}
The Crab Nebula has become a standard candle in TeV astronomy; it has
been detected by many groups and its integral flux appears constant.
(We refer to the ``steady'' emission, not emission modulated at the
pulsar period.  We have not detected the latter at TeV energies.  See
paper by G.~Gillanders et al., in these proceedings.) It is also well
on the way to becoming a standard candle with regard to TeV spectral
content.  Synchrotron emission from the Crab covers a remarkably broad
range terminating at about 10$^8$ eV, where a new component attributed
to inverse Compton scattering begins. It is this component that we
detect.  The TeV spectrum is sensitive to the primary electron
spectrum, the nebular magnetic field and the spatial distribution of
electrons and magnetic field within the nebula.

In this paper we briefly describe two methods for extracting TeV
spectra, compare results from the Whipple Observatory Imaging
Cherenkov Telescope for the 1988/89 and 1995/96 observing seasons and
comment on implications for the physics of the nebula.  The methods,
developed at Iowa State University and the University of Leeds, are
based on Monte Carlo simulations using completely independent code and
use different approaches in determining the overall gain of the
Cherenkov telescope.  In the ISU approach, the gains of the
photomultiplier tubes, mirror reflectivities, etc., are measured and
combined to find the overall gain.  In the Leeds approach, the
observed brightness distribution of cosmic-ray images is combined with
cosmic-ray simulation results to determine the overall gain of the
telescope.  The methods are described in detail in ``Paper I,''
Mohanty et al., (1997), and the resulting TeV spectrum is put into
context of other observations with implications for the physics of the
nebula in ``Paper II,'' Hillas et al., (1997).  The latter paper also
compares our Crab TeV spectrum with those of other groups.

\vfill\eject
\section{2. METHOD 1: A TRADITIONAL APPROACH}
A straightforward approach to the determination of TeV spectra was
developed at Iowa State University.  Three components are required.
First, a method of distinguishing gamma-ray images from background
cosmic-ray images.  The standard method is to use ``supercuts'' as
described in, e.g., Punch et al. (1996).  The images are characterized
by second order moments giving the {\it width}, {\it length}, {\it
distance} of the image centroid from the optic axis and {\it alpha},
the angle by which the image major axis misses passing through the
optic axis.  More that 99\% of the background can be rejected by
requiring small values of {\it width}, {\it length} and {\it alpha}.
However, this procedure results in a strong bias against the images of
higher energy gamma rays which tend to be longer and broader and hence
more cosmic-ray like.  Mohanty (1995) has modified the procedure so
that the image selection criteria depend on the total brightness or
{\it size} of the image as well.  (The {\it size} can be used as an
estimate of the energy.)  The telescope collection areas for simulated
gamma rays for the 1995/96 season using standard and ``extended''
supercuts is shown in Fig.~1.

\begin{wrapfigure}[19]{r}{9.5cm}
\epsfig{file=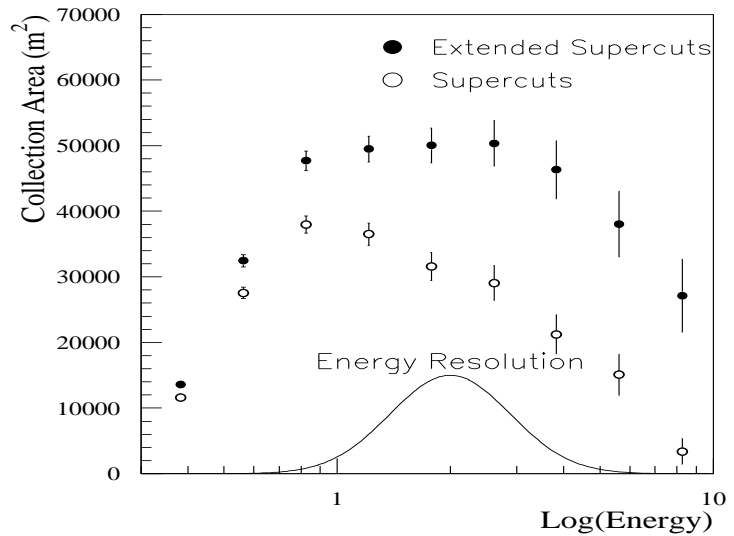,bbllx=43truept,bblly=51truept,bburx=260truept,bbury=344truept,width=7.0cm,height=7.0cm,angle=0}
\caption{\it The collection areas for standard and extended supercuts
for the 1995/96 observing season are shown above.  The resolution
function is also plotted.}
\end{wrapfigure} 

The second component needed is a way to estimate the energy of each
gamma-ray image.  Two desirable criteria are (a) good resolution and
(b) negligible bias.  The former is important to detect small
structures in the spectrum and the latter is important to avoid
distortions.  We obtained a resolution of $\Delta E/E \sim 0.36$ with
negligible bias by using a second order polynomial in {\it size} and
{\it distance} as described in Paper I.  The energy resolution
function is, to a good approximation, Gaussian in the variable
log($E$).  It is plotted at an arbitrary energy in Fig.~1.

In data taken for spectral analysis, each on-source observation is
followed by an ``off-source'' observation covering the same range of
range of azimuth and elevation angles.  The images for both types of
observation are selected for gamma-like events.  The estimated
energies from corresponding observations are histogrammed and the
difference ascribed to gamma rays from the source.  This can then be
fit by a power law or the fluxes extracted as described in Paper I.

\section{3. METHOD 2: AN INTUITIVE APPROACH}
A different approach with emphasis on verifiability was developed at
the University of Leeds.  There are two pieces to this approach, a
method for selecting images likely to have been initiated by cosmic
gamma rays and a method for determining the primary gamma-ray energy
spectrum from the observed {\it size} spectra.  Earlier descriptions
of this approach are in (Hillas and West, 1991) and (West, 1994).

The selection criterion is a ``cluster'' or ``spherical'' method in
which a single parameter is used to characterize the gamma-ray-like
nature of an image and correlations between image parameters are
incorporated naturally.  Simulated gamma rays produce images with four
parameters ({\it width}, {\it length}, {\it distance} and {\it alpha})
that populate a four-dimensional space. Each real image can be tested
to see if it is likely to be a gamma-ray image by the value of the
Mahalonobis distance between it and the centroid of the cluster.
(This is equivalent to scaling and rotation of the axes so that the
window is spherical and fluctuations uncorrelated.)  The selection
window is defined by the expected position and dimensions of the
gamma-ray parameter cluster for several broad ranges of image {\it
size.}

In order to extract a spectrum, a {\it size} histogram is then
computed for on-source and off-source observations and the difference
histogram is ascribed to gamma rays.  A simulated {\it size} spectrum
is then computed starting with a power-law primary energy spectrum for
gamma rays.  A weight is given to each simulated gamma ray and, by
adjusting these weights, the spectrum can be varied so that its {\it
size} spectrum matches that of the difference histogram.  This method
is simple, easy to implement and avoids the complexity of calculating
collection areas and bias-free energy estimates.  It is described in
detail in Paper I.

\section{4. RESULTS}
\begin{wrapfigure}[23]{r}{9.5cm}
\epsfig{file=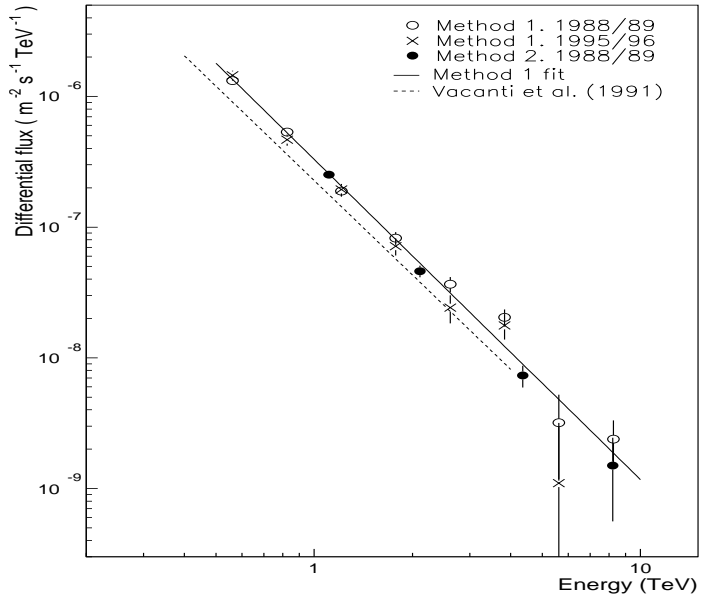,width=7.5cm,height=7.0cm,bbllx=0truept,bblly=30truept,bburx=420truept,bbury=600truept,angle=0}
\caption{\it The Crab spectrum in the range 0.3 to about 8 TeV
extracted using Methods 1 (open circles) and 2 (solid circles) for the
Whipple 1988/89 database and using Method 1 (x's) for the 1995/96
database are shown above.  Also shown is a fit to the combined Method
1 results (solid line) as well as an earlier spectrum (dashed line)
taken from Vacanti et al. (1991).}
\end{wrapfigure} 
The spectra obtained from the 1988/89 and 1995/96 seasons using Method
1 and 1988/89 season using Method 2 are in good agreement as shown in
Fig.~2.  As described in Paper I, we have tested the sensitivity of
the results to uncertainties in the Monte Carlo simulations and have
found that the results are relatively robust.  The combined TeV fluxes
from both seasons from Method 1 are well fit with the simple power law
spectrum in which the differential flux ($J(E)$) is given by:
\begin{equation}
(3.3 \pm .2 \pm .7)\,10^{-7}\,(\frac{E}{\hbox{TeV}})^{-2.45\pm.08\pm.05}
\end{equation}
in units of $\hbox{m}^{-2}\,\hbox{s}^{-1}\,\hbox{TeV}^{-1}$ where the
first errors are statistical and the second are our estimate of
systematic errors.

However, the simple extrapolation of this fit to lower energies yields
fluxes far in excess of those observed by EGRET as is clear from
Fig.~3 which shows a quadratic fit of log($J$) vs.\ log($E$) to our
data and to an averaged point representing the EGRET flux (Nolan et
al., 1993) at 2 GeV.  This fit may be written:
\begin{equation}
J(E) = (3.25)\,10^{-7}\,(E/\hbox{TeV})^{-2.44-0.135\hbox{log}_{10}(E)}
\hbox{m}^{-2}\,\hbox{s}^{-1}\,\hbox{TeV}^{-1}.
\end{equation}

\section{5. COMMENTS ON INTERPRETATION}
Most of the early inverse Compton models for TeV gamma rays assumed a
constant magnetic field in the principal source region where these are
produced (e.g., Gould 1965 or Rieke and Weekes 1969) whereas more
recent models (De Jager and Harding, 1992 or Aharonian and Atoyan)
incorporate hydrodynamic plasma/field flow making the calculations
more complex and the results probably more realistic.  Here, we try to
stay close to the data and make the simpler assumption that the field
is constant.  The broad synchrotron emission band apparently extends
up to 10$^{8}$ eV and is boosted to higher energies via inverse
Compton scattering.  The scattering giving rise to TeV gamma rays
occurs in the Klein-Nishina rather than in the Thomson scattering
regime.  This implies that the electrons giving rise to our detected
gamma rays must have energies in the range of 2-10 TeV and the
corresponding scattered photons would mostly have energies of about
0.005 to 0.3 eV.  This conclusion is only very weakly model dependent
(see Paper II).  Since electrons with energies of a few TeV generate
synchrotron radiation at about 0.4 keV in a field of about 25 nT (see
next paragraph), the Einstein Observatory X-ray pictures of Harnden
and Seward (1984) also show the part of the nebula emitting TeV gamma
rays.

\begin{wrapfigure}[24]{r}{9.5cm}
\epsfig{file=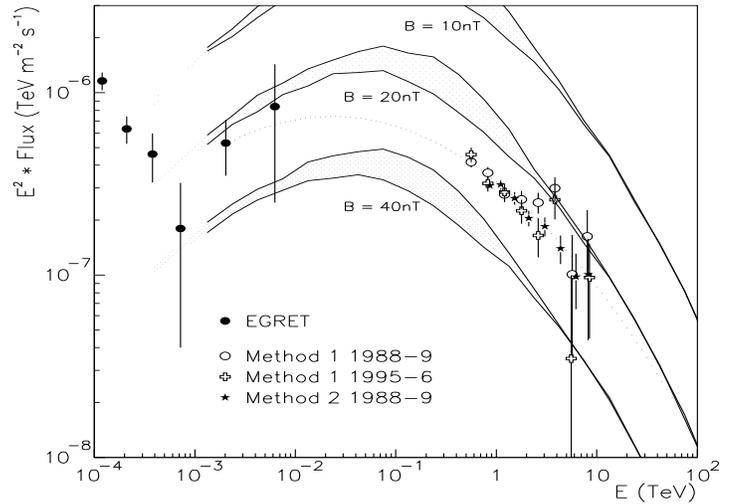,width=7.5cm,height=7.0cm,bbllx=0truept,bblly=0truept,bburx=450truept,bbury=550truept,angle=0}
\caption{\it Whipple and EGRET observations of the Crab gamma-ray
spectrum.  The dotted line is a fit to the WHipple points and a single
2 GeV flux value.  Full-line curves are predicted inverse Compton
fluxes for 3 differenct assumed B fields.}
\end{wrapfigure} 

For assumed magnetic field values and the observed synchrotron flux,
it is possible to deduce the spectrum of primary electrons, presumably
generated in the shock at the termination of the pulsar wind (see,
e.g., Coroniti and Kennel 1985). From the ambient photon density and
deduced electron spectrum, the TeV flux can be calculated, and results
for B fields of 10, 20 and 40 nT are shown in Fig.~3.  The shaded
regions reflect uncertainties arising from ill-defined UV to
soft-X-ray region of the synchrotron photon spectrum.  As can be seen
from the figure, the effective B field must lie between 20 and 40 nT
with 27 nT falling very near to our TeV data.  Since even more
energetic electrons only keep their energy for a short time, they
should exist only near the pulsar wind shock.  Hence, measurement of
the TeV spectrum over a wider energy range may probe spatial
variations in the nebular magnetic field (see Paper II).

\section{6. ACKNOWLEDGEMENT}
This work is supported by grants from the US DOE and NASA, by PPARC
in the UK and by Forbairt in Ireland. 

\section{7. REFERENCES}
\setlength{\parindent}{-5mm}
\begin{list}{}{\topsep 0pt \partopsep 0pt \itemsep 0pt \leftmargin 5mm
\parsep 0pt \itemindent -5mm}
\vspace{-15pt}
\frenchspacing

\item Aharonian, F.A. and Atoyan A.M, Astropart. Phys. {\bf 3}, 275
(1995) 

\item Atoyan, A.M. and Aharonian, F.A., MNRAS {\bf 278} 525
(1996). 

\item Coroniti, F.V. and Kennel, C.F.,  in ``The Crab Nebula and
related supernova remnants,'' ed. M.C.Kafatos and R.B.C. Henry, CUP, p
25 (1985). 

\item De Jager, O.C., and Harding, A.K., Ap. J., {\bf 396}, 161 (1992). 

\item Gould, R.J., Phys. Rev. Lett., {\bf 15}, 577 (1965). 

\item Harnden, F.R.,Jr. and Seward, F.D., Ap. J. {\bf 283} 279 (1984). 

\item Hillas, A.M. and West, M., Proc. 22nd ICRC Dublin {\bf 6} 472 (1991).

\item Hillas, A.M., et al, ``Paper II,'' to be submitted to
Ap. J. (1997).

\item Mohanty, G., PhD Thesis Iowa State University (1995).

\item Mohanty, G., et al., ``Paper I,'' Astroparticle Physics, 
submitted (1997).

\item Nolan, P.L. et al., Ap. J., {\bf 409}, 697 (1993). 

\item Punch, M., et al., Proc. 22nd ICRC Dublin {\bf 1} 464 (1991).

\item Rieke, G.H., and Weekes, T.C., Ap. J., {\bf 155}. 429 (1969). 

\item Vacanti, G., et al., Ap. J. {\bf 377}, 467 (1991).

\item West, M., PhD Thesis University of Leeds (1994).

\end{list}

\end{document}